# Magneto-transport behaviour of $Bi_2Se_{3-x}Te_x$: Role of disorder

**E P Amaladass[1,3], T R Devidas[1], Shilpam Sharma[1], C S Sundar[1], A Bharathi[2] and Awadhesh Mani[1,3]**

[1]Materials Science Group, IGCAR, Kalpakkam 603102
[2]UGC-DAE CSR Node, Kokilamedu 603104

[3]**Corresponding Author**: edward@igcar.gov.in, mani@igcar.gov.in

## Abstract

Magneto-resistance and Hall resistance measurements have been carried out in fast-cooled single crystals of $Bi_2Se_{3-x}Te_x$ (x = 0 to 2) in 4 - 300 K temperature range, under magnetic fields up to 15 T. The variation of resistivity with temperature that points to a metallic behaviour in $Bi_2Se_3$, shows an up-turn at low temperatures in the Te doped samples. Magneto-resistance measurements in $Bi_2Se_3$ show clear signatures of Shubnikov – de Hass oscillations that gets suppressed in the Te doped samples. In the $Bi_2SeTe_2$ sample, the magneto-resistance shows a cusp like positive magneto-resistance at low magnetic fields and low temperatures, a feature associated with weak anti-localisation (WAL), that crosses over to negative magneto-resistance at higher fields. The qualitatively different magneto-transport behaviour seen in $Bi_2SeTe_2$ as compared to $Bi_2Se_3$ is rationalised in terms of the disorder, through an estimate of the carrier density, carrier mobility and an analysis in terms of the Ioffe-Regel criterion with support from Hall Effect measurements.

## 1. Introduction

Topological insulators (TI) are a new class of materials in which band inversion in the bulk band structure, gives rise to a metallic state at the surface, that has a unique Dirac dispersion with its spin locked to its momentum [1-4]. Following the initial elegant experiments using angle resolved photoemission spectroscopy (ARPES) [5-8] and scanning tunnelling microscopy (STM) [9] that provided an unambiguous evidence for the presence of Dirac dispersed 2D states, the current interest in this field has shifted to exploring the novel transport properties of 2D surface states (SS). Observation of quantum phenomenon in magneto-resistance (MR) such as Shubnikov-de Hass oscillation (SdH) and weak anti-localisation (WAL), a positive cusp at the vicinity of zero field, are unique signatures of the SS in TI. However, the true nature of the SS is often masked by the parallel bulk conductance from excess carriers due to anti-site disorders and vacancies [10, 11]. In addition, different defects levels in surface and bulk can lead to the formation of two dimensional electron gas (2DEG) at the bulk/surface interface due to band bending [12] and mimic the 2D surface states.

Previous reports showed that Se vacancies ($V_{Se}$) and $Se_{Bi}$ anti-site disorders are the predominant donor-type defects in $Bi_2Se_3$ system in all growth conditions, whereas in $Bi_2Te_3$, p-type anti-site disorders $Te_{Bi}/Bi_{Te}$, are energetically more favourable as compared to the vacancies [11]. Therefore, alloying these end compounds compensates the bulk defect states and result in a bulk insulating state as it is observed in $Bi_2SeTe_2$ ternary [13]. Our earlier studies in $Bi_2Se_3$ with excess Se, had clearly indicated that, stacking faults which has formed due to excess Se seem favourable in the manifestation of the surface state in the transport behaviour [14].

In the present study, we investigate the magneto-resistance behaviour of disordered Te substituted $Bi_2Se_3$ along with Hall Effect measurements. It is shown that the metallic behaviour seen in an undoped $Bi_2Se_3$ changes to an insulating behaviour with Te doping. In addition, the SdH oscillations seen in the undoped sample disappears with Te doping. Excess Te composition sample, $Bi_2SeTe_2$ exhibits a WAL to weak localization (WL) crossover with field and a non-linear slope in Hall Effect measurement at 4.2 K. The results of the present studies are rationalized in terms of electronic disorder introduced by Te doping which is quantified by the Ioffe-Regel parameter ($k_f l$), mobility and carrier density and their evolution with temperature.

## 2. Experimental details

Single crystals of $Bi_2Se_{3-x}Te_x$ (with x=0, 1, 1.8 and 2) were grown using high purity (99.999%) Bi, Se and Te powders as starting materials. To accomplish the growth, stoichiometric mixtures of elements were melted at 850˚C for 6 hours in an evacuated sealed quartz tube environment, followed by rapid cooling at the rate of 137$^{o}$C/hour. This step of rapid cooling is drastically different from the other reports in the literature [13, 15] and it is intended to introduce disorder in the system. Synchrotron X-ray diffraction measurements were carried out on the powdered single crystal samples at room temperature, at the beam line BL-12 of the Indus-2 synchrotron in Indore, India, using 12. 6 keV X-rays. The powdered crystals were placed in a circular depression made on Kapton tape. The data was collected using a MAR3450 image plate detector in the transmission geometry. Fit2D program was employed to convert the 2D image from the image plate detector to 1D Intensity versus 2θ plot. In addition, single crystals were characterized by Laue diffraction measurements, carried out in transmission and reflection geometry using molybdenum X-ray source and HD-CR-35 NDT image plate system.

Magneto-transport and Hall measurements were carried out in both, the van der Pauw geometry and six-probe linear geometry in a commercial, 15 T Cryogen-free system from Cryogenic Ltd., UK. For the transport measurements, the cleaved crystal was pasted on a puck holder using double sided tape. Electrical contacts were made using a 20 micron Au wires with room temperature-cured silver paste. With a special switching and scanning unit from Keithley Instruments, both resistivity and Hall Effect measurements were carried out on each of the single crystals, simultaneously. Repeated experiments on randomly selected crystals showed similar behaviour providing support for the results presented in this paper.

## 3. Results

### 3.1. X-ray Diffraction

The XRD patterns on the powdered single crystals obtained from the synchrotron source, for samples with nominal compositions of $Bi_2Se_3$ (BS), $Bi_2Se_2Te$ (BST), $Bi_2Se_{1.2}Te_{1.8}$ and $Bi_2SeTe_2$ (BTS) are shown in figure 1(a). All the diffraction peaks could be indexed to Rhombohedral structure, and the variation of lattice parameters with Te content is shown in figure 1(b). It is seen that both the *a*- and *c*-lattice parameters increase monotonically with the increase in the Te composition as expected, since the larger Te atom replaces a smaller Se atom, and the variation of lattice parameters is in accordance with similar reported studies [16]. The broadening of the (015) diffraction peak with Te substitution as shown in the right panel of figure 1(a) is indicative of increasing disorder in the sample. This is also reflected in Laue diffraction patterns acquired in back scattering geometry as observed in figure 1(c). A clear six-fold symmetry with a line of close spots that is observed in the undoped crystals is seen to get reduced and distorted with increasing Te substitution and become less distinctive in the high concentration Te doped specimen indicative of the increasing crystalline disorder in the sample with Te substitution.

### 3.2. Resistivity and Hall Measurements vs Temperature

Figure 2(a) shows the resistivity (ρ) versus temperature (T) behaviour of $Bi_2Se_{3-x}Te_x$ (x=0, 1, 1.8 and 2) samples. It is evident from the figure that the pristine $Bi_2Se_3$ crystal shows a metallic behaviour with a discernible saturation in resistivity below 30 K. Such a saturation tendency at low temperature has been reported to stem from the excess carriers from Se vacancies in the sample [15]. In the experiments by Ren et al. [17] using different preparation procedure a weak upturn in resistivity at low temperature is reported. In the $Bi_2Se_{1.2}Te_{1.8}$ and BTS samples, the positive temperature coefficient of resistivity at high temperatures changes to a negative temperature co-efficient as the temperature is lowered.

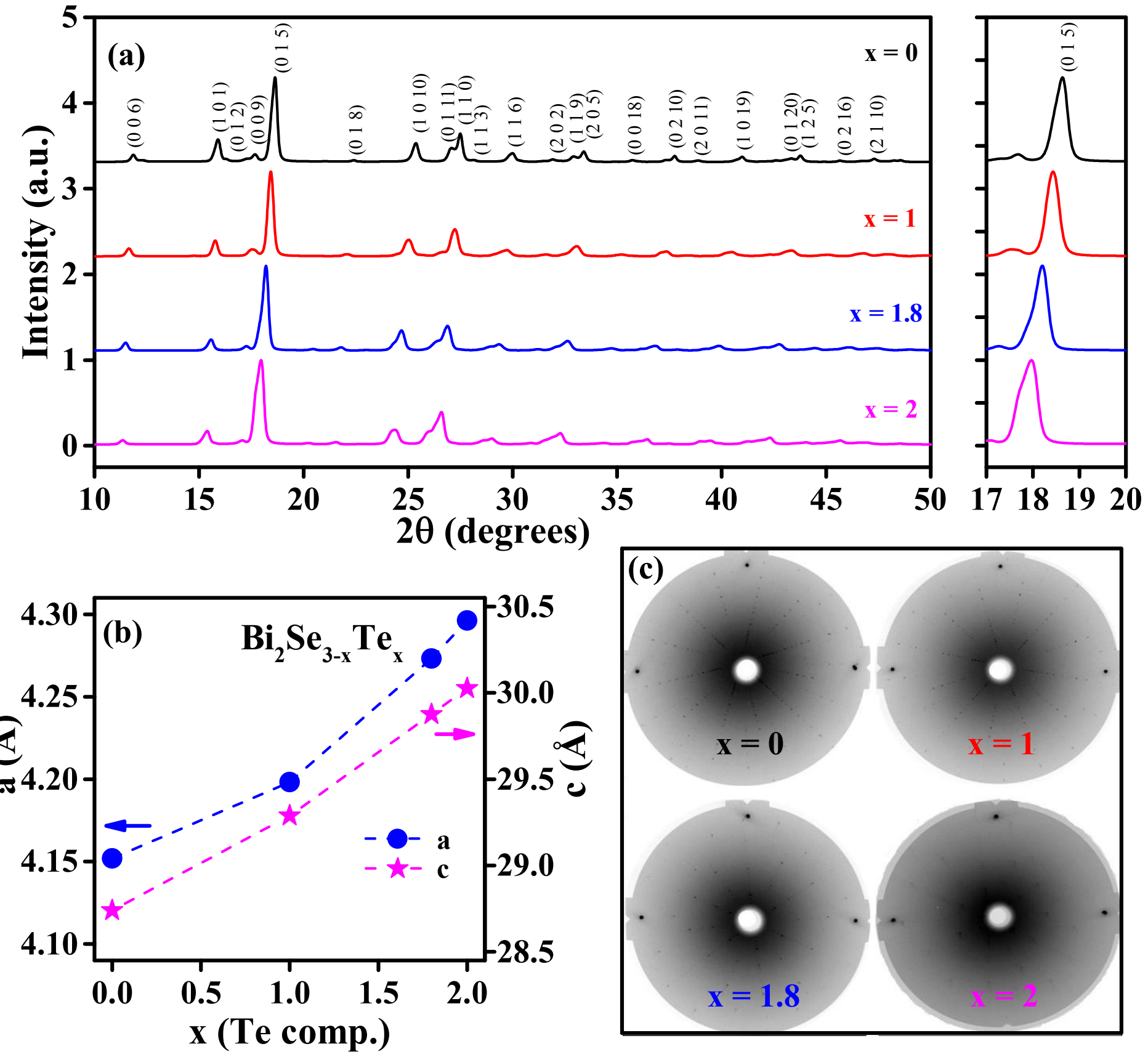

Figure 1(a). Synchrotron powder XRD patterns of $Bi_2Se_{3-x}Te_x$ samples. The graphs has been shifted vertically for clarity, the indices of the Rhombohedral structure are shown. Right panel of (a) shows the broadening of (015) peak upon Te substitution. (b) The variation of *a*- and *c*- lattice parameters obtained from the XRD patterns. (c) Laue diffraction patterns in the backward scattering geometry shows the six fold symmetry with distinct spot in Se rich samples (x=0,1) and the gradual degradation of crystalline order inferred from the distortion and elongation of spots along the radial direction upon Te substitution (x=1.8,2).

The temperature dependence of resistivity ρ(T) and its order of magnitude are different from that reported by Ren et al. [13, 15] for BTS sample where resistivity changed by two orders of magnitude and clearly showed activation behaviour at high temperature and a 3D variable range hopping (VRH) at low temperature. In our experiments, the low temperature upturn of high Te concentration sample was best fitted with Mott VRH type behaviour for temperature range 30 -130 K, as shown in figure 2(b). The data could also be fitted with Arrhenius activation behaviour for similar temperature range as shown for BTS sample in the inset of figure 2(b) and the activation energy was found to be 1.5 meV. Since the temperature range for fit is too narrow, we are not in a position to unequivocally establish the mechanism of transport (VRH/ Arrhenius) from the present data. It is seen from the inset of figure 2(a) that the resistivity at 300 K increases with Te content, a behaviour that can arise due to a reduction in the carrier concentration and/or a suppression of the carrier mobility.

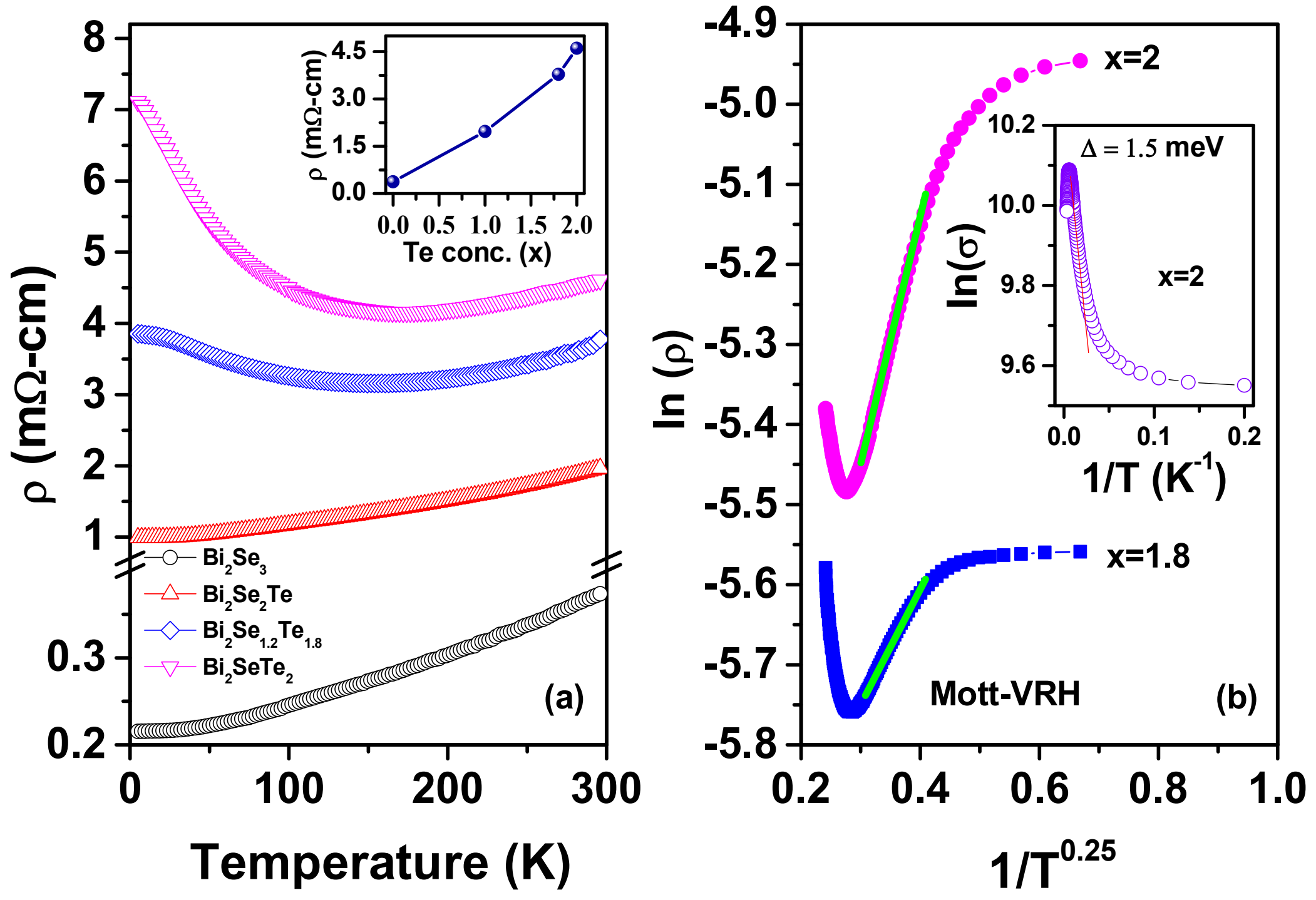

Figure 2. (a) The variation of resistivity as a function of temperature of $Bi_2Se_{3-x}Te_x$ samples. A metal to insulator type of behaviour is evident upon Te substitution; Inset shows an increase in the resistivity at 300 K with Te concentration. The right panel (b) shows the VRH fit for high Te concentration (x=1.8, 2) samples and the inset shows the Arhenius fit for x=2 sample.

In order to understand the reason for the change in resistivity behaviour with Te addition, Hall Effect measurements were carried out. Figure 3 shows the Hall resistance as a function of magnetic field at various temperatures in the pristine and Te doped samples. In the pristine $Bi_2Se_3$ sample, Hall resistance is linear with magnetic field and is nearly temperature independent (see figure 3 (a)). This indicates that transport is determined by a single carrier (electron type) and the lack of temperature dependence implies that carrier density is unaffected by temperature. A similar behaviour of the Hall resistivity is also seen in $Bi_2Se_2Te$ sample. In the higher Te concentration samples, viz., $Bi_2Se_{1.2}Te_{1.8}$ and $Bi_2SeTe_2$ (see figure 3 (c)-(d)), the Hall resistance shows a non-linear behaviour with respect to applied magnetic field at low T, suggestive of presence of transport from two types of carriers [13].

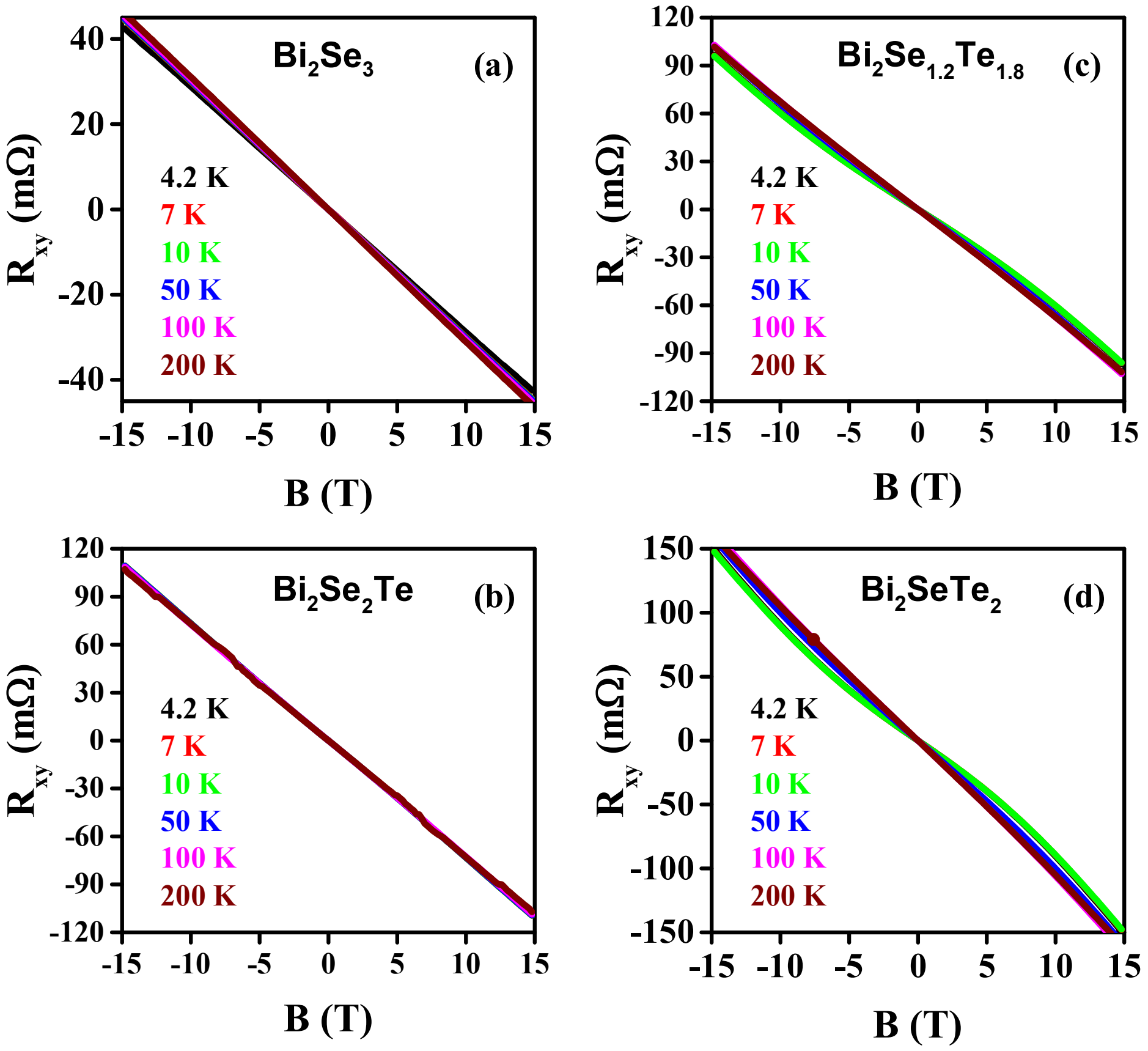

Figure 3. The variation of Hall resistance as a function of magnetic field at various temperatures in $Bi_2Se_3$ and Te doped samples showing a linear behaviour for low Te concentration (a, b) and a non-linear behaviour at low T for higher Te concentration (c, d).

Interestingly, as discussed in Sec. 3.3, this difference in the variation of Hall resistance with Te doping is linked to the corresponding magneto-resistance behaviour. From the resistivity and Hall Effect measurements (cf. figure 2 and 3), the carrier concentration (n) and the mobility (μ) of carriers are evaluated using the following relations:

$$n = 1 / (R_H\, e)$$

$$\mu = 1/\rho n e$$

Where $R_H$ is Hall coefficient in $m^3C^{-1}$, e is the charge of electron in coulombs, ρ is resistivity in ohm-cm. Further, from the resistivity and the carrier concentration we evaluate the Ioffe-Regel parameter [12] given by $k_f l = (3\pi)^{2/3}\left(\frac{1}{(e^2 \rho n^{1/3})}\right)$ where $k_f$ is the Fermi wave vector and l is the electron mean free path. The variation of carrier density (n), mobility (μ) and $k_f l$ as a function of temperature is shown in figure 4 and their variation as a function of Te content at 4.2 K is shown in the right panel.

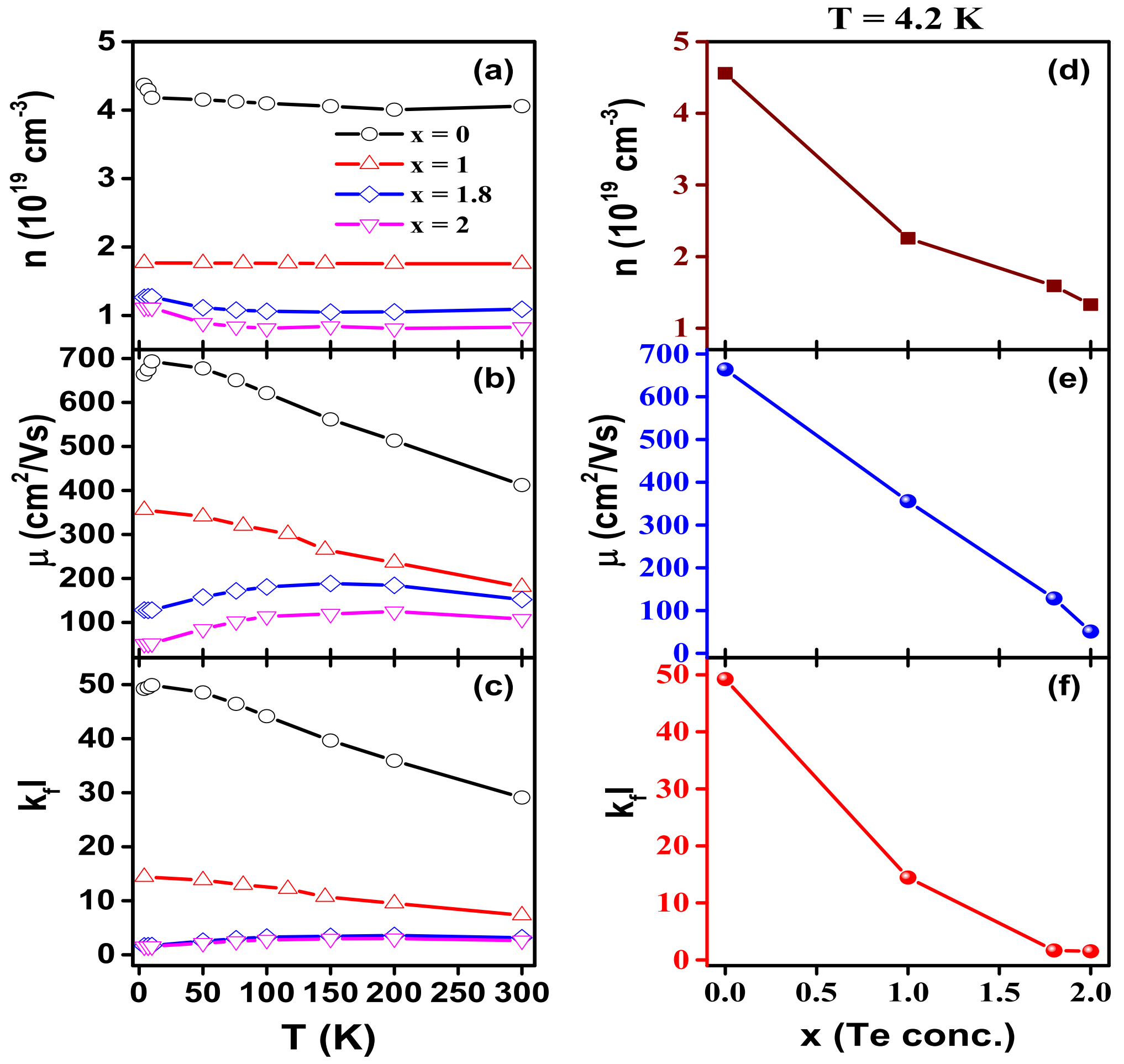

Figure 4. Temperature and Te concentration dependence of carrier concentration (n), mobility(μ) and the Ioeffe-Regel parameter ($k_f l$), as obtained from the analysis of Hall and resistivity data.

A close inspection of figure 4 brings out following important features: (i) The carrier concentration decreases with increase in Te content, and is largely temperature independent. The observed decrease in carrier density with Te doping is contrary to the expectation as the band gap decreases [23] from 0.3 eV in $Bi_2Se_3$ to 0.17 eV in $Bi_2Te_3$. (ii) The Hall mobility, increases with decrease in temperature in the low Te samples (x=0,1), the degree of increase in mobility with decrease in temperature, slows down with increase in Te content, whereas for large Te content (x=1.8,2) the mobility decreases with decrease in temperature. This decrease in Hall Mobility ($\mu = e\tau/m$) with increase in Te concentration implies a decrease in scattering time$\tau$, and hence implies an increase in disorder with increase in Te. (3) The $k_f l$ values obtained from the resistivity and Hall data, decreases with Te content and becomes close to 1 in the high concentration Te samples, particularly at low temperatures (see figure 4 (f)). This implies that the electronic disorder induced by Te doping leads to the localisation of carriers which can account for the appearance of upturn in ρ(T) of  high Te doped $Bi_2Se_{3-x}Te_x$ (x=1.8, 2) as seen in figure 2(a).

### 3.3. Magneto-transport measurements (SdH Oscillations and WAL/ WL behaviour)

In order to see the effect of the drastic change in mobility upon Te substitution on magneto-transport, magneto-resistance (R(B)-R(0)) measurement have been carried out on $Bi_2Se_{3-x}Te_x$ (x=0, 1, 1.8 and 2) samples at 4.2 K in the field range of -15 T to 15 T and is shown in figure 5 (a). The variation of magneto-resistance (MR) and Hall resistance as a function of magnetic field (B) is shown figure 5 (a) - (b) respectively. Pristine $Bi_2Se_3$ sample shows a positive MR with discernible Shubnikov-de Haas (SdH) oscillations for B > 7 T. Fourier transform of the oscillations in $Bi_2Se_3$ and $Bi_2Se_2Te$ are shown in the left and right insets of figure 5 (a) respectively.

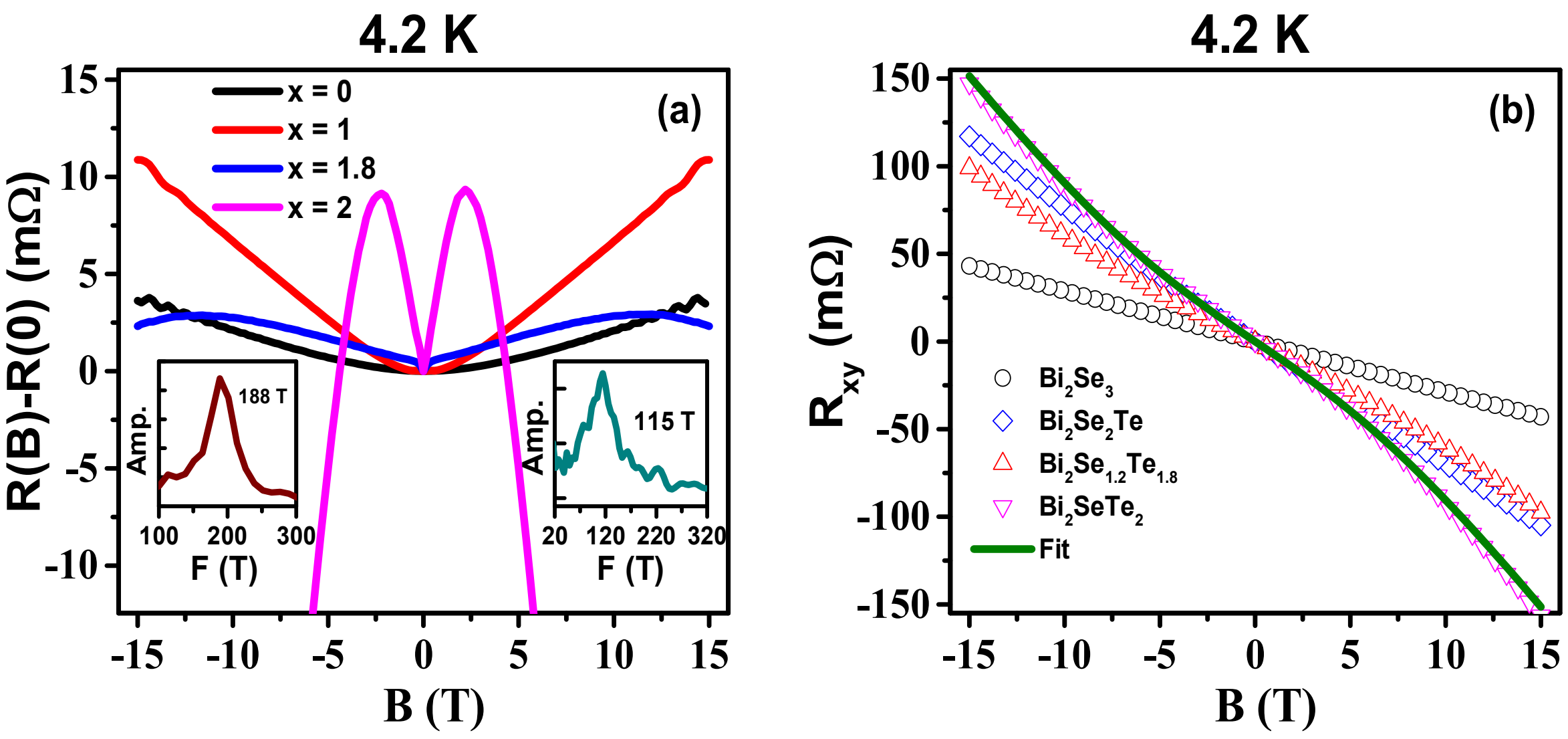

Figure 5 (a) Variation of magneto-resistance with magnetic field measured with magnetic field parallel to the c-axis of the crystal at 4.2 K, in the $Bi_2Se_{3-x}Te_x$ series. With increase in Te concentration negative magneto-resistance emerges. Left and right insets show the Fourier transform of the R(B) curves of $Bi_2Se_3$ and $Bi_2Se_2Te$, respectively. (b) The transverse resistance as a function of magnetic field in the $Bi_2Se_{3-x}Te_x$ series. With increase in Te concentration the $R_{xy}$ plots become non-linear.

The oscillation frequency, thus deduced, is found to decrease from 188 T for $Bi_2Se_3$ to 115 T for $Bi_2Se_2Te$. This implies contraction of the Fermi surface with Te substitution [18]. It is also noted that the oscillations are not at a single frequency for $Bi_2Se_2Te$ unlike $Bi_2Se_3$. This points to lack of coherence of the Landau levels due to disorder caused by Te doping. The SdH oscillation vanishes for $Bi_2Se_{1.8}Te_{1.2}$ and $Bi_2SeTe_2$ samples, and interestingly the MR at low temperatures shows a low field cusp behaviour that is associated with Weak anti-localisation (WAL) [19]. The MR of x=1.8 shows a positive slope up to ~10 T and changes to a negative slope for B > 10 T. The x= 2 sample shows more pronounced positive cusp for -2T< B < 2 T, reaches a maximum at 2 T, and then starts to decrease on increasing the field similar to WL. It has been shown from earlier studies [19, 20] that the cusp like positive magneto-resistance behaviour attributed to WAL gives way to negative magneto-resistance on account of opening of band gap at the Dirac point due to magnetic impurity [21], or due to hybridisation of the bottom and top surface states in very thin samples [18, 22]. As our sample is neither thin film nor has any magnetic impurities, we could rule out such interpretation of the WAL-WL crossover.

To obtain insight into the observed MR results, Hall measurements on the same samples are displayed in figure 5 (b). It is seen that Hall resistance versus magnetic field is linear, which implies that transport occurs due to a single type of carrier in samples (x=0,1). In contrast, the larger Te substituted samples (x=1.8, 2) shows a non-linearity in the Hall versus magnetic field plot which points to the two carrier behaviour. Interestingly, the emergence of two carrier transports in figure 5 (b), correlates with the presence of transition from positive to negative magneto-resistance in figure 5 (a). Such non-linear behaviour has already been observed in similar TI samples and is reported to be due to the parallel contribution of both surface and bulk carrier to the transport [13, 15, 17, 23]. Therefore to account for the mixed surface and bulk carriers, the Hall data of BTS sample (x=2) is fitted with the standard two band model using following equation [13].

$$\rho_{xy} = -\frac{(R_s\rho_n^2 + R_n\rho_s^2)B + R_sR_n(R_s + R_n)B^3}{(\rho_s + \rho_n)^2 + (R_s + R_n)^2B^2}$$

Where $R_n$, $R_s$ are the Hall coefficient of bulk and surface electrons respectively $\rho_n$ and $\rho_s$ are their resistivity of bulk and surface electrons respectively. This fitting as shown in figure 5(b) yield bulk carrier concentration $n_n = 4.8 \times 10^{20}$ cm$^{-3}$ and surface carrier concentration $n_s = 1 \times 10^{12}$ cm$^{-2}$ along with the bulk mobility $\mu_n = 54 \times 10^{-6}$ cm$^2$/V.s and surface mobility $\mu_s = 0.06$ cm$^2$/V.s. It is seen that the surface mobility is higher than that of the bulk.

Figure 6 (a)-(d) show the temperature evolution of the MR for all the four samples. For the $Bi_2Se_3$ sample, the oscillatory magneto-resistance behaviour is seen for temperatures up to 10 K and gets suppressed as the temperature is increased. The oscillatory part of MR rides on a linear high field MR behaviour, whose magnitude decreases with increase in temperature. In the case of $Bi_2Se_2Te$, (cf. figure 6(b)), the MR behaviour is qualitatively similar to $Bi_2Se_3$, albeit with a substantially large MR. However, the MR of x=1.8 and 2 shown in figures 6 (c)–(d) exhibit WAL behaviour which is qualitatively different from that of $Bi_2Se_3$. In x=1.8 sample the negative slope seen at B > 10 T vanishes when temperature exceed 10 K. Whereas, in x=2 sample a systematic transition upon increasing temperature is seen. From 4.2 K to 10 K the WAL cusp decreases and vanishes at 50 K. On further increasing the temperature, T > 50 K, a transition to positive MR is observed. Interestingly the Hall data shown in figures 3 (c)–(d) also shows a transformation from a non-linear behaviour at low temperature (T <10 K) to a linear behaviour on increasing T. The non-linearity in Hall data is more prominent in BTS sample. Whereas the Hall data of x= 0, 1 samples remained linear for all the temperature range except for slight slope change. Since both WAL and WL are quantum phenomena they are not observed at higher temperatures, where a regular positive magneto-resistance is seen.

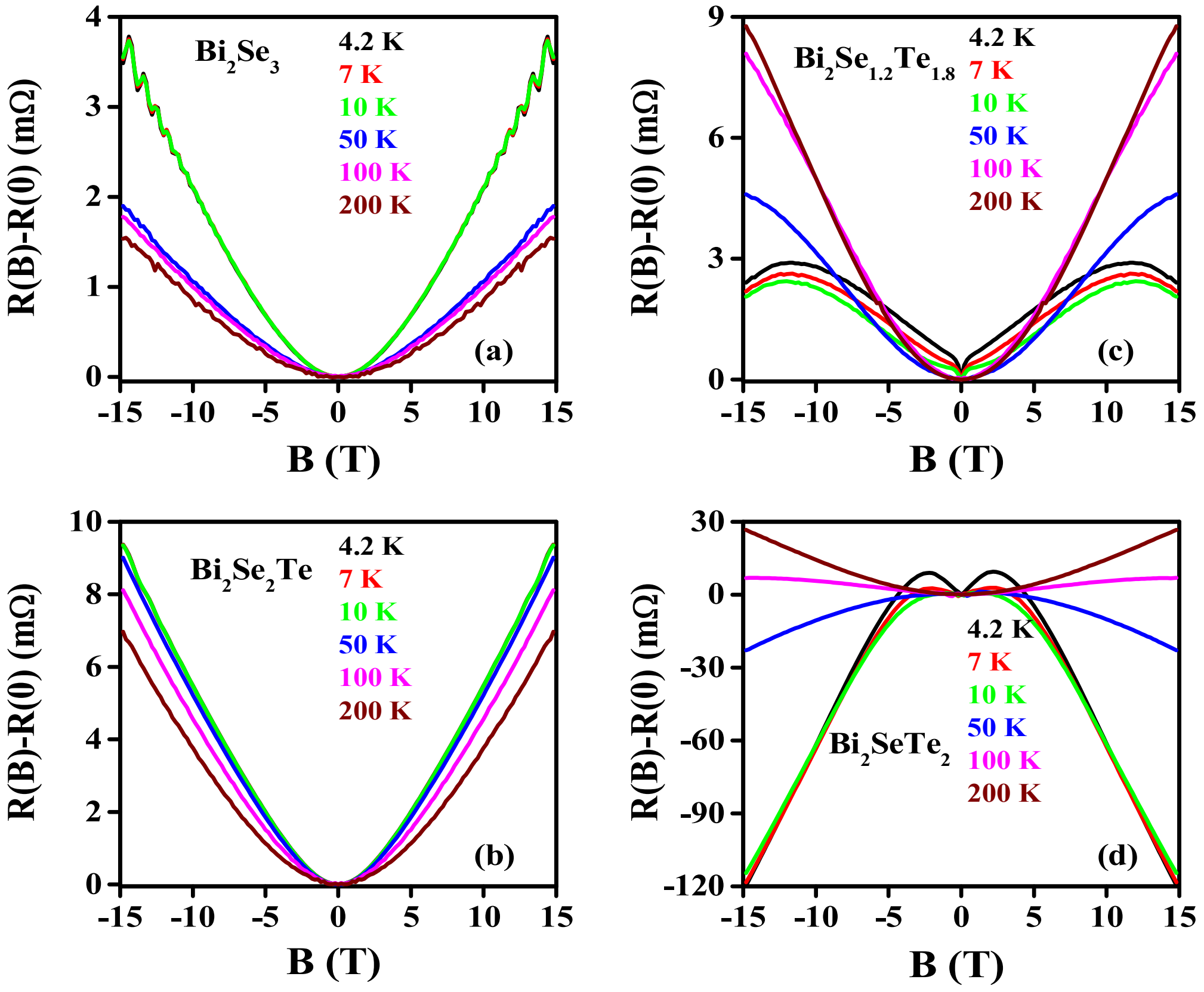

Figure 6. The variation of magneto-resistance as a function of magnetic field at various temperatures in (a) $Bi_2Se_3$, (b) $Bi_2Se_2Te$ (c) $Bi_2Se_{1.2}Te_{1.8}$ and (d) $Bi_2SeTe_2$. SdH oscillation observed in $Bi_2Se_3$ at low T vanishes on increasing T. Qualitatively different MR behaviour is seen in high Te samples. Negative MR is evident in $Bi_2Se_{1.2}Te_{1.8}$ and $Bi_2SeTe_2$ at B> 10T and B > 2 T, respectively. Upon increasing temperature the negative MR gradually transform to positive MR.

## 4. Summary and Conclusions

Electronic transport studies were carried out on $Bi_2Se_{3-x}Te_x$, (x= 0 to 2) topological insulator single crystals as a function of magnetic field up to 15 T and in temperature range of 4.2 – 300 K. While SdH oscillation is seen in MR for pristine and low Te doped $Bi_2Se_3$ samples, an interesting feature exhibiting evolution of WAL behaviour is observed upon higher Te doped $Bi_2Se_{3-x}Te_x$ samples. The WAL is found to undergo WL transition with increase of magnetic field. Interestingly this coincides with the emergence of the two carrier transport as seen in the Hall measurements. First, we note that both the SdH and WAL behaviour are quantum manifestations of magneto-transport. In the $Bi_2Se_3$ sample, the mobility is high (See figure 4) and this is conducive to the observation of SdH oscillations. The addition of Te introduces an electronic disorder in $Bi_2Se_{3-x}Te_x$ systems, which in turn reduces their electronic mobility and the mean free path (See figure 4). The decrease in the mean free path ($l_e$) makes it conducive for the condition $l_\varphi > l_e$ ($l_\varphi$ : phase coherence length ) to be met leading to the quantum corrections to the transport, viz., the WAL behaviour that points to the involvement of topological surface state in the transport behaviour [12]. With the increase of magnetic field, this WAL feature is seen to transform to the WL behaviour – as has been observed in other experiments [18, 24]. As seen in figure 6, the change from WAL to WL occurs when the Hall resistance shows a non-linear behaviour, that requires analysis in terms of two carrier transport involving both the surface and bulk electrons [19]. The significant result from the present experiments is that there is a change from SdH to WAL behaviour with Te doping, which has been attributed to the electronic disorder. This is one of the few experiments where SdH to WAL transition is seen through variation of disorder. Also we believe that electronically manipulating the Fermi level in BTS sample by gating experiments at low temperatures will allow us to control the surface/bulk contribution, which in turn will shed more light on the WL and WAL phenomenon in a disordered TI system.

**Acknowledgements**

We are grateful to the Dr. G. Amarendra, scientist in-charge for facilitating the use the magneto-resistance facility at the UGC-DAE CSR node at Kalpakkam. One of the authors, T.R. Devidas gratefully acknowledges the Department of Atomic Energy for the Senior Research Fellowship. We acknowledge Ms. Anitha K for helping us in sample preparation, Dr. A.K. Sinha for XRD beam time at BL-12 of the Indus-2 synchrotron in Indore, India. We also acknowledge Dr. A Thamizhavel and Dr. N. Subramanian for Laue measurements at TIFR, Mumbai. Dr. A. Bharathi acknowledges the support of CSIR Emeritus Fellowship and Dr. C. S. Sundar acknowledges the DST for J. C. Bose Fellowship.